\documentclass[a4paper,twoside]{article}

\usepackage{epsfig}
\usepackage{subcaption}
\usepackage{calc}
\usepackage{amssymb}
\usepackage{amstext}
\usepackage{amsmath}
\usepackage{amsthm}
\usepackage{multicol}
\usepackage{pslatex}
\usepackage{algorithmic}
\usepackage{algorithm}
\usepackage{caption}
\usepackage{subcaption}
\usepackage{array}
\usepackage{apalike}
\usepackage[bottom]{footmisc}
\usepackage{SCITEPRESS}     % Please add other packages that you may need BEFORE the SCITEPRESS.sty package.

\begin{document}

\title{Towards An Online Incremental Approach to  Predict Students Performance}

\author{\authorname{Chahrazed Labba and Anne Boyer}
\affiliation{\sup{} Loria, CNRS, University of Lorraine, Vandœuvre-lès-Nancy, France}}
%\affiliation{\sup{2}Department of Computing, Main University, MySecondTown, MyCountry}
%\email{\{chahrazed.labba, anne.boyer\}@loria.fr}

\abstract{Analytical models developed in offline settings with pre-prepared data are typically used to predict students' performance. However, when data are available over time,  this learning method is not suitable anymore. Online learning is increasingly used to update the online models from stream data. A rehearsal technique is typically used, which entails re-training the model on a small training set that is updated each time new data is received.  
  The main challenge in this regard is the construction of the training set with appropriate data samples to maintain good model performance. Typically, a random selection of samples is made, which can deteriorate the model's performance. In this paper, we propose a memory-based online incremental learning approach for updating an online classifier that predicts student performance using stream data. The approach is based on the use of the genetic algorithm heuristic while respecting the memory space constraints as well as the balance of class labels. In contrast to random selection, our approach improves the stability of the analytical model by promoting diversity when creating the training set. As a proof of concept, we applied it to the open dataset OULAD. Our approach achieves a notable improvement in model accuracy, with an enhancement of nearly 10\% compared to the current state-of-the-art, while maintaining a relatively low standard deviation in accuracy, ranging from 1\% to 2.1\%.}
\keywords{Machine learning, Genetic Algorithm, Classification, Prediction, Online Learning}

%\ccsdesc[500]{Tracing Learning and Teaching}
%\ccsdesc[300]{Finding evidence of learning}
%\ccsdesc{Prediction}
%\ccsdesc[100]{Learning}



\maketitle

\section{Introduction}
One of the primary concerns in e-learning environments is the identification of the students  who are experiencing learning difficulties. Conventionally, analytical models developed in offline settings using pre-prepared data are usually applied to predict students' performance. The majority of these models operate in batch mode, reading and processing all of the training data with the strong assumption that the data is static and available in advance. Data availability in the context of e-learning is highly dependent on students' interaction with the learning content, which is affected by a variety of external factors (deadlines, mental health, mood...). As a result, gathering all relevant training samples at once is impossible. Thus, as data become available over time, traditional methods of training and evaluating analytical models become obsolete and unsuitable.

To address this challenge, online incremental learning is increasingly being used to update online Machine Learning (ML) models with new data received over time. According to the existing definitions \cite{gepperth2016incremental,yang2019survey}, incremental learning is the process of learning from streaming data with limited memory resources while maintaining a good model performance. However, training the model on stream data only, may result in a deterioration in model performance due to what we call catastrophic forgetting \cite{hayes2020remind}. %In fact, unlike humans who have the ability to preserve old knowledge even when new knowledge is acquired, ML models eventually forget the old knowledge once trained on a new one. 
The first solution that comes to mind to address this issue is to retrain the model on the old data. However, due to privacy and/or memory constraints \cite{chang2021student}, access to old data cannot be assured in the context of online incremental learning. 

%\begin{figure}
    %\centering
    %\includegraphics[width=0.9\linewidth]{figures/random.png}
    %\caption{Variation in Accuracy over 38 Weeks (10 Runs) using the random selection strategy proposed in \cite{labba2022} with OULAD dataset to predict students performance}
    %\label{fig1}
%\end{figure}

In the literature \cite{rebuffi2017icarl,kirkpatrick2017overcoming}, we distinguish mainly three methods to meet this challenge including memory-based, regularization-based and network-architecture-based approaches. This paper is concerned with memory-based approaches that consist in using rehearsal techniques to recall a small episodic memory from  previous tasks (data) when training on the new tasks, thereby reducing loss on previous tasks. %Overall, the use of incremental learning is more developed, especially for image classification. Several memory-based approaches \cite{rebuffi2017icarl,yan2021framework}  have been proposed to build and update exemplar sets to retrain models in incremental learning. However, this methods are not well suited to the students performance predictions. This is due to the fact that the context and the type of data are different. %In image classification, for example, the exemplar set is incrementally updated; each time a new class label is detected. Whereas for old data samples for an existing class label, the only possible update is the deletion of some samples because a new class label has been added and memory space cannot be exceeded. However, in our case, this is no longer valid because the observations in the old data change over time. Thus, each time new data is received, the exemplar set is updated to include the selection of new observations for new class labels as well as new observations for old class labels.
%Another challenge that is less addressed in incremental learning is the model frequency updates. In fact, the frequency with which model updates are performed is strongly related to the knowledge encoded in newly received data and its relevance to model learning.

To the best of our knowledge, \cite{labba2022} is the sole work that has addressed and evaluated a memory-based incremental learning strategy for predicting students' performance. Their approach introduces an episodic memory designed to maintain a balanced training set of both old and new data, updating the model when necessary. Assume that the size of the training set is fixed at 100 samples and that at a time $t$, it contains two class labels: failure and success. The data set is well balanced between these labels. As time passes at \(t+1\), new data arrive, which may include updates of previous observations or entirely new observations. The task is to incorporate a selection mechanism that randomly chooses samples from the new data received at t+1 to update and refresh the training set established at t. This approach ensures that the training set remains up-to-date and reflects the most recent data without exceeding the 100-sample limit. However, a notable limitation arises from the random selection of training samples to build the training set (called also exemplar set), potentially leading to a less model performance due to the inclusion of poorly selected samples. %The Fig.\ref{fig1} shows 10 executions of the approach proposed in \cite{labba2022}, to predict weekly student performance using the OULAD dataset (10 runs for each week). We evaluated the model (random forest), in terms of accuracy, by re-training it on different training sets constructed each week using a random selection of training samples. The Fig.\ref{fig1} shows the variations in the model's accuracy due to randomness, where outliers (red dots) are almost observed for every week over the 10 runs, and standard deviation values vary from 1\% to 9\%. 

To limit the impact of randomness, we propose a new memory-based incremental learning strategy to predict students' performance that is based on the use of the Genetic Algorithm (GA) to build the training set. The adoption of a genetic algorithm  is not arbitrary.  In this work, we believe that an initial population composed of many individuals \footnote{an individual represents a set of samples selected for the training}, combined with a guided strategy for their generation, enables the creation of a more diverse group of individuals. This in turn contributes to improving the stability and robustness of the approach to build the training exemplar. In contrast, the random selection process of training samples generates only one random solution at a time, which limits its ability to promote diversity and may result in increased variation in terms of model performance.  

To prove the efficiency of our approach, a comparison with \cite{labba2022}, using OULAD, is performed. Overall, for the random selection strategy \cite{labba2022}, we observe the widest range of standard deviation in accuracy, extending from 1\% to 9\%. Conversely, our GA-based approach shows significantly lower variations, limited to a range of 1\% to a maximum of 2.1\%. Further, we proved it's more effective to consider loss as a score for the GA fitness function  to select the training samples, as this resulted in reduced overall loss while maintaining an acceptable measure of accuracy.  

The rest of the paper is organized as follows: Section \ref{related} presents the related work. Section \ref{approach} describes the online incremental process based on the GA to build the exemplar set. Before concluding in Section \ref{conclusion}, Section \ref{experiments} presents the experimental results using the OULAD dataset. 

\section{Related Work}\label{related}

Overall, the use of incremental learning is more developed, for image classification. In \cite{rebuffi2017icarl}, the authors presented iCaRL, a training approach that enables incremental learning of classes from stream data. The iCaRL strategy relies on small training sets (called exemplar sets) for the observed classes, which can be updated as new ones emerge. %A fixed amount of memory is set aside for all the set of examplers. When a new class label appears in the stream data, a new examplar set is created, and the old ones are updated by removing samples to meet the memory use condition. The sample selection for inclusion in a new exemplar or removal from an older one is not random and is based on an approximation of the average feature vector. %In other words, the samples are prioritized in order to create a representative set of training samples. The method has been used only to classify images. Even though this approach ensures incremental class learning, it is not well suited to classification problems where observations for old classes change and must be updated over time. 
In \cite{yan2021framework}, the authors presented an online learning strategy for semantic segmentation that allows learning new visual concepts on a continuous basis for pixel-wise semantic labeling of images. The strategy involves a re-labeling approach for augmenting annotations as well as an efficient rehearsal-based model adjustment with dynamic data sampling to overcome the catastrophic forgetting. The later one is achieved by using a replay buffer to save samples for previous tasks and use them to regularize the learning each time new tasks are received. %Despite the fact that the proposed algorithm for building the training buffer is based on a random sample selection, it takes into account the class-balance. 
In \cite{he2020incremental}, the authors proposed an incremental learning framework to overcome the problem of catastrophic forgetting when learning new classes and the problem of data distribution over time referred as concept drift. The framework was tested to classify images using the CIFAR-100 and ImageNet-1000 datasets. In \cite{sirshar2021incremental} the authors presented a novel framework that can incrementally learn to identify various chest abnormalities by using few training data.
the framework is based on an incremental learning loss function that infers Bayesian theory to recognize structural and semantic inter-dependencies between incrementally learned knowledge representations. %In \cite{losing2018incremental}, the authors compared eight popular incremental methods representing different algorithm classes using stationary and non-stationary datasets. A set of metrics including the accuracy, the robustness and the error classification rate are used to assess the algorithms. 

When it comes to predicting student performance incrementally, most of the research is oriented towards the comparison of incremental algorithms. In \cite{kulkarni2014prediction}, the authors compared four classifiers that can run incrementally. The aim is to recommend the suitable algorithm to use in assessing students performance within an incremental learning context.  In \cite{ade2014instance}, the authors compared three approaches of incremental learning to determine the suitable way to handle students stream data. The used approaches include instance-based, batch-based and ensembling of instance-based incremental learning. In \cite{kotsiantis2010combinational}, the authors proposed an incremental learning technique that combines an incremental version of Naive Bayes, the 1-NN and the WINNOW algorithms. The aim is to predict the student's performance within a distance education environment by using incremental ensemble based on a voting methodology. To the best of our knowledge, \cite{labba2022}  is the only work in which  a memory-based incremental learning strategy to predict students' performance was tackled and evaluated. The proposed approach addresses the frequency of updating an online model. An episodic memory that stores a balanced exemplar set of old and new data to train the model when necessary is used. An algorithm is proposed to continuously update the exemplar set as long as stream data is received. However, the main issue with the proposed solution is that the exemplar set is built using a random selection of the training samples. As a result, we may end up with a set of examples that, rather than maintaining a good model performance, degrades it due to  poor samples that might be selected using a random strategy. 
\section{GA to Build a training Exemplar} \label{approach}

Due to privacy concerns and/or memory limitations, access to old data cannot be guaranteed in the context of incremental learning \cite{chang2021student} . Thus, to overcome the common problems (e.g catastrophic forgetting), memory-based techniques that involve utilizing a small episodic memory from prior data when training on the new data is used. The main idea consists in maintaining a balanced exemplar training set of both old and new data in order to re-train the model when necessary while using a fixed amount of memory. Newly received data is no longer used alone to train the model; instead, an exemplar set containing both old and new data is used. In \cite{labba2022}, the exemplar set is updated each time new data are received. The update includes the add of observations for new classes and the update of the observations of old classes. However, the main issue is that the exemplar set is built using a random selection of the training samples. However, this random selection may result in a poor model performance.  The problem of selecting new and old observations to build the training exemplar set  is NP-Hard. Given a collection of old and new observations, the optimal training exemplar is a set or subset of these samples. This is a discrete selection method. It is very expensive to determine the optimal set of observations with a permutation of possibilities, especially with the strong memory constraints. To find the best set, we propose to use a genetic algorithm to generate many alternatives when constructing the exemplar training set and then selecting the one that improves model performance in terms of accuracy and loss while meeting memory constraints.

The genetic algorithms use an evolutionary approach that consists in the following phases: For sample selection, the first step is to generate a population based on subsets of the possible samples while respecting the exemplar size.  From this population, the subsets are used as training data to train a model. Each time a model is trained on a subset, it is evaluated using a fitness function which can be accuracy or loss. Once each member of the population is evaluated against the fitness function, a tournament is conducted to determine which subsets will continue into the next generation. The next generation is composed of the winners of the tournament, with some crossover and mutation. Crossover consists of updating the winning sample sets with samples from the other winners, while mutation introduces or removes some samples at random. 

\subsection{Representation of individuals}
The first step is to determine how each individual will be represented. Since the goal is to find, at a given time $t_i$, the best subset of data observations from both old and new observations to train a model, each individual must be a representation of the subset of observations that it holds. Individuals will be represented as a string of length equal to the number of both old and new observations at $t_i$, with each char of the string corresponding to a training sample and a value of 1 or 0 depending on whether the sample is selected or not. Each individual will thus occupy n bytes, where n is the initial number of old and new observations at $t_i$ , and m is the maximum number of samples to be activated within an individual ($m \ll n$, m corresponds to the size of the exemplar).

\subsection{Generate population}

The next step is to generate a population of a certain size that will serve as the algorithm's starting point. Individual generation can be completely random or slightly controlled. In our work, the number of samples to consider for each class label can be controlled. Indeed, a random generation of individuals can result in the construction of unbalanced exemplar sets in terms of class representation. However, since the main goal is to find the best subset of samples that improves the model's performance, the generation cannot be completely random. To respect memory constraints and consider a balanced representation among the class labels, it is interesting to force the individual generator to generate individuals with a fixed number of 1s per class label. As a result, all individuals will have a fixed number of active samples while considering the balance criterion as well as the memory limitations. 

The Algorithm \ref{algpopulation} allows the generation of individuals by considering both random and controlled generation. The control is performed on the number of samples per class label to ensure balance. While random generation is applied when selecting samples in the same class label. The algorithm takes as input the old and new observations represented by $(D,Y)$, the size of the population to be generated $size_p$, and the size of the exemplar set $E\_size$. It provides as output the population $\mathcal{P}$ that contains the generated individuals. The algorithm starts by initializing the list $\mathcal{P}$ (Line 1). Then, it determines the set of class labels presented in the input data $(D,Y)$ and computes the number of samples to be selected for each class label in the exemplar set (Line 2 - Line 8). For each generated individual in $\mathcal{P}$, the algorithm selects a set of samples to be used in training at random for each class label, by assigning a value of $1$ if the sample is included and $0$ otherwise (Line 10 - Line 29). Once the samples per class label are selected, the algorithm verifies if the final generated individual exists or not in $\mathcal{P}$ (Line 24 - Line 28). If this is the case, the process of generation will be repeated. Else the generated individual will be added to the list $\mathcal{P}$. The entire process is repeated until the required number of individuals is reached.
\begin{algorithm}[t]
\caption{Generate a population }
\label{algpopulation}
\begin{algorithmic} [1]
\REQUIRE $(D,Y),size\_p, E\_size$
\ENSURE $\mathcal{P}$ 
\STATE $\mathcal{P} \leftarrow list()$
\STATE $class\_labels \leftarrow get\_true\_lables(Y) $
\STATE $size\_class\_labels \leftarrow list()$
\FOR{($c \in class\_labels $)}
\STATE $size\_label \leftarrow get\_size\_samples((D,Y),c)$
\STATE $size\_class\_labels\leftarrow add(size\_label)$
\ENDFOR
\STATE $ratio=E\_size \// size(class\_labels)$
\FOR{($i \in size\_p $)}
\WHILE{True}
\STATE $individual \leftarrow ""$
\FOR{$j \in size\_class\_labels$}
\STATE $sub\_individual \leftarrow ""$
\FOR{($char \in j $)}
\STATE $count\_1=get\_number\_of\_1()$
\IF{$count\_1== ratio$}
\STATE $concat(sub\_individual,0)$
\ELSE
\STATE $concat(sub\_individual,rand(0,1))$
\ENDIF
\ENDFOR
\STATE $concat(individual,sub\_individual)$
\ENDFOR
\IF{$individual \in \mathcal{P}$}
\STATE $Write(Repeat\_generation\_process)$
\ELSE
\STATE $Break$
\ENDIF
\ENDWHILE
\STATE $\mathcal{P} \leftarrow concat(\mathcal{P},individual) $
\ENDFOR
\end{algorithmic}
\end{algorithm}
\subsection{Evaluate each individual in the population}
For each of the generated individuals in $\mathcal{P}$,  a model will be trained  with a dataset containing only the active samples for that individual. After training the model on a subset of samples from each individual, we must assess its performance. In this work, accuracy and loss are used as metrics to quantify the effectiveness of the trained models using different exemplar sets. The Algorithm \ref{algeval} presents the way the fitness function is defined to evaluate the performance of the individuals (exemplar sets). The algorithm takes as input the old and new observations represented by $(D,Y)$, a small test dataset $(X_{test}, Y_{test})$, a model $\mathcal{M}$ to be trained and evaluated, the list of generated individuals $\mathcal{P}$ and the score type $S_{type}$ (accuracy or loss) to evaluate the individual. It provides as output, a dictionary $\mathcal{W}$ that assigns to each individual the corresponding accuracy and loss. For each individual in $\mathcal{P}$,the algorithm first copies the parameters of the model to be trained into a new one (Line 3). Then, it retrieves the training data that correspond to the active samples for the individual under evaluation (Line 4). The model is trained on these data, and the test data is used to make predictions. According to the score type, the model is assessed either in terms of accuracy or loss (Line 5 - Line 11). Once completed, the individual and the associated evaluation metric are added to $\mathcal{W}$. 

\begin{algorithm}[t]
\caption{Evaluate the Population in terms of accuracy or loss }
\label{algeval}
\begin{algorithmic} [1]
\REQUIRE $(D,Y),(X_{test}, Y_{test}),\mathcal{P},\mathcal{M}, S_{type}$
\ENSURE $\mathcal{W}$ 
\STATE $\mathcal{W} \leftarrow \left\{ \right\}$
\FOR{$individual \in \mathcal{P}$}
\STATE $\mathcal{M}_{0} \leftarrow copy(\mathcal{M}) $
\STATE $X_{train}, Y_{train} \leftarrow get\_train\_data()$
\STATE $\mathcal{M}_{0}.fit(X_{train},Y_{train})$
\STATE $predictions \leftarrow \mathcal{M}_{0}.predict(X_{test}) $
\IF{$S_{type}=="Accuracy"$}
\STATE $S= compute\_accuracy(predictions, Y_{test})$
\ELSE
\STATE $S= compute\_loss(predictions, Y_{test})$ 
\ENDIF
\STATE $\mathcal{W} \leftarrow put(individual, S) $
\ENDFOR
\end{algorithmic}
\end{algorithm}

\begin{algorithm}[t]
\caption{the reproduction process : crossover and mutation }
\label{algrepro}
\begin{algorithmic} [1]
\REQUIRE $p_1, p_2, E\_size, (D,Y) $
\ENSURE $child_1,child_2$ 
\STATE $nb\_samples=length(individual_1)$
\STATE $crosspoint=Random(nb\_samples)$
\STATE $child_1=p_1[:crosspoint]+p_2[crosspoint:]$
\STATE $child_2=p_2[:crosspoint]+p_1[crosspoint:]$
\STATE $child_1=Mutate(child_1,nb\_samples,(D,Y),E\_size)$
\STATE $child_2=Mutate(child_2,nb\_samples,(D,Y),E\_size)$
\end{algorithmic}
\end{algorithm}

\begin{algorithm}[h]
\caption{Mutate }
\label{algcrossmut}
\begin{algorithmic} [1]
\REQUIRE $child,nb\_samples,(D,Y),E\_size, prob$
\ENSURE $new\_child$ 
\STATE $list\_index=list()$
\STATE $S=E\_size/length(class\_labels)$
\STATE $new\_child=""$
\STATE $class\_labels \leftarrow get\_true\_lables(Y) $
\STATE $size\_class\_labels \leftarrow list()$
\FOR{($c \in class\_labels $)}
\STATE $index\_start\_end=get\_indexes((D,Y),c)$
\STATE $list\_index= add(list\_index,index\_start\_end)$
\ENDFOR
\FOR{$index \in list\_index$}
\STATE $sub\_child=child[index[0]:index[1]]$
\FOR{$i, char \in sub\_child $}
\STATE $ran=Random()$
\IF{$ran <prob$}
\IF{$char== 0 \textbf{and} count(1,sub\_child)<S$}
\STATE $new\_char=1$
\STATE $sub\_child=sub\_child[:i]+new\_char+sub\_child[i+1:]$
\ELSIF{$char== 1$}
\STATE $new\_char=0$
\STATE $sub\_child=sub\_child[:i]+new\_char+sub\_child[i+1:]$
\ENDIF
\ENDIF
\ENDFOR
\STATE $new\_child=concat(new\_childsub\_child)$
\ENDFOR
\end{algorithmic}
\end{algorithm}
\subsection{Reproduce individuals: selection, crossover and mutation}
Following the computation of each individual's scores (accuracy and loss), a selection is performed to determine which individuals will continue to the next generation. For genetic algorithms there are different techniques to perform the selection phase such as the use of the roulette wheel selection, the stochastic universal sampling and the tournament selection. In our work, we consider a fitness based selection approach where the fittest $k$ individuals are selected to participate in the generation phase. The individuals with the best scoring metric will be selected to reproduce and pass their genes to the next generation. Then, the most important stage of a genetic algorithm is crossover. A crossover point is chosen at random from the genes for each pair of individuals (parents) that will be mated. Offspring are produced by exchanging genes between parents until the crossover point is reached. Some of their genes may be subject to mutation with a low random probability in some of the newly formed offspring. As a result, some of the activated and not activated samples may be reversed. The goal of mutation is to increase population diversity and prevent early convergence. The Algorithm \ref{algrepro} introduces the reproduction process once the selection is performed. The algorithm takes as input the two parents $p_1$ and $p_2$ to be mated, the data observations $(D,Y)$ and the exemplar size $E\_size$. It provides as output two individuals $child_1$ and $child_2$. The algorithm starts by generating randomly a cross point (Line 1 -Line 2). Then, it mates $p_1$ and $p_2$ and returns $child_1$ (Line 3) and $child_2$ (Line 4). Once the children are created, the mutation phase begins. The Algorithm \ref{algcrossmut} introduces the mutation process. The algorithm takes as input the child to be mutated $child$, the length of the individual $nb\_samples$, the data observations $(D,Y)$, the exemplar size $E\_size$ and the probability of mutation $prob$. It provides as output  the mutated child $new\_child$. The algorithm starts by determining for each class label in the data $(D,Y)$, the start and end indices of the samples that represent it (Line 4 - Line 9). The mutation is then applied by class label. Indeed, when the crossover is performed, the child's size constraint may be violated, and we may end up with a child with a greater number of activated samples than is required. In addition, due to the randomness associated with the generation of the crossover point, the data equilibrium  criteria in terms of class label may also be violated. For each class label (line 10 - Line 25), the algorithm starts by copying the sequence of bits that represents this class label in the child to mutate (Line 11). Then, according to the generated random value, if it is lower than the probability value $prob$, the algorithm checks if the character to be mutated is equal to $0$ and if the number of $1$ in the sequence is lower than the representation ratio of the class label (Line 14 - Line 15). If this is the case, the algorithm reverses the bit and the sequence is updated (line 16 -17). Otherwise, if the character is equal to 1, then the mutation is applied and the sequence is updated (Line 18 - Line 20). 
The new sequence is then concatenated to the mutated child $new\_child$ (Line 24). In this way, we ensure that the reproduction process provides solutions that respect memory constraints (exemplar set size) as well as class-label balance criteria.  

\subsection{Main loop of Genetic Algorithm}
To generate a set of appropriate training examples, all of the preceding steps should be carried out in the following order. The first step is to generate the initial population. In fact, the number of individuals to generate is an input parameter. Once generated, each individual is evaluated using one of the defined fitness functions, including accuracy and loss. When one of the individuals reaches the optimal value for the accuracy or loss functions when calculating the fitness score, the genetic algorithm is interrupted and the individual in question is provided as the best solution. For this work, the optimum value for accuracy is set to 100\%, while the loss is set to 0. These values can be always modified as required.  

If non of the individuals reaches the goal state, the selection process to select the best parents is performed. Then, the reproducing algorithm including crossover and mutation is invoked. At the end of this process a new population is generated and then evaluated. This process is repeated n times, where n is an input parameter and represents the number of iterations. During the whole execution process, the individual with the best score value is kept as the best solution.

\section{Experimental Results} \label{experiments}
\subsection{Case Study}

To validate our GA-based incremental learning process, we used the Open University Learning Analytics dataset (OULAD)\cite{kuzilek2017open}. 
This dataset is a collection of data from the Open University's Virtual Learning Environment (VLE). The dataset contains information about student activities and interactions with the VLE, such as accessing course materials, participating in online discussions, and submitting assignments. The OULAD dataset has been widely used in learning analytics research and has been made available for academic use. Overall, it presents the data related to seven courses \footnote{['AAA', 'BBB', 'CCC', 'DDD', 'EEE', 'FFF', 'GGG']}, where we have more than 30000 enrolled students. In our work, we are interested in predicting student performance as early as possible in time. To accomplish this, we defined a set of features for the dataset OULAD. We distinguish the following indicators \cite{labba2022}\cite{ben2021depth} : 1) Demographic data; 2) Performance that denotes the submitted exams and the grades; 3) Engagement that describes the learner interaction with the VLE content; 4) Regularity that denotes the progress made by the learner in terms of achieved VLE activities and the number of submitted exams and 5) Reactivity that is described by the time taken to submit an exam as well as the time between successive connections to the VLE.

We adopt a weekly-based prediction strategy. The problem is formalized as a binary classification problem. The classification consists of two classes: failure and pass.
On each week $w_i$, a student is defined by a tuple $X=(f_1, ..,f_m, y)$ where $f_1, ..,f_m$ are the features and $y$ is the class label. The prediction of the student's class label is made based on their final true label.
The aim is to predict students at risk of failure as early as possible while taking into account the progressive availability of data over time. 

The experimental section presents the results of the prediction against the AAA course over the year 2013. Overall, we have 363 enrolled student and 39 weeks \footnote{The original data are provided by day, the data were processed to calculate the features on a weekly basis}. 

\subsection{Experiments}

We compared the GA-based incremental learning process using both loss and accuracy as score types to the incremental process based on a random selection introduced in \cite{labba2022}. 

We used a small forgetting value (0.01) to invoke model retrain at each week and check how model performance varies over 39 weeks in terms of two metrics including accuracy and loss. The maximum size of the  training exemplar is set to 80. for the rest of this section the displayed  results represent the average value over 10 runs for each week, for both accuracy and loss metrics.
When configuring the GA, we fixed the initial population size to 20 individuals and  the number of iterations to 5. 

\subsubsection{The GA-based incremental process: Stability over runs}

\begin{figure}
    \centering
    \includegraphics[width=0.85\linewidth]{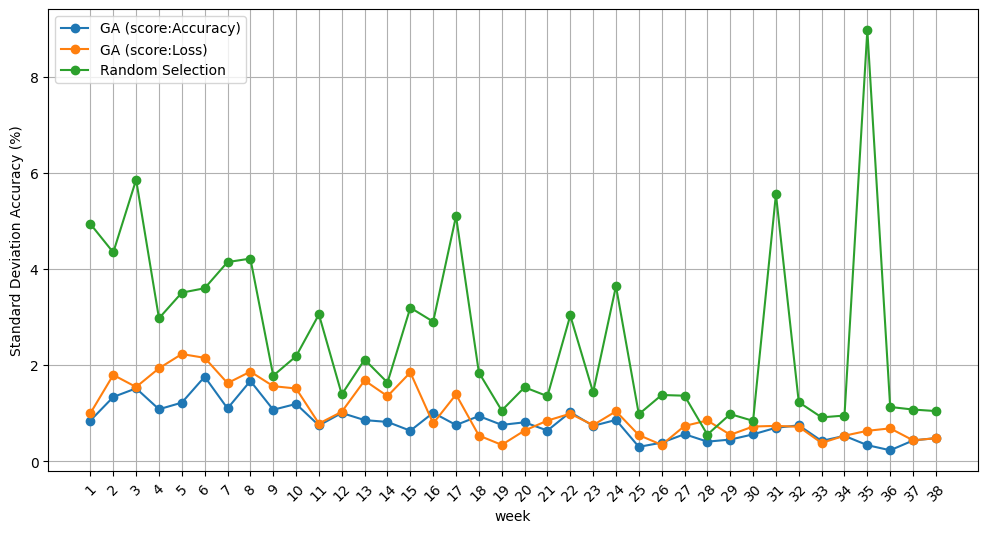}
    \caption{Standard deviation in terms of accuracy over 10 runs: GA Accuracy vs GA Loss vs Random Selection}
    \label{std}
\end{figure}

Our objective is to mitigate the influence of random sample selection in \cite{labba2022} when constructing the exemplar training set for the purpose of retraining the online analytical model. As a result, our goal is to minimize the standard deviation in model accuracy when retraining it using the constructed exemplar. The Fig.\ref{std} illustrates the weekly standard deviation of accuracy, with each week's standard deviation calculated from a set of 10 runs. For the random selection strategy \cite{labba2022}, we observe the widest range of standard deviation in accuracy, extending from 1\% to 9\%. Conversely, our GA-based approach shows significantly lower variations, limited to a range of 1\% to a maximum of 2.1\%. In terms of stability, the adoption of a genetic algorithm with an initial population of 20 individuals, combined with a guided strategy for their generation, enables the creation of a more diverse group of individuals. This in turn contributes to improving the stability and robustness of our approach. In contrast, the random selection process generates only one random solution at a time, which limits its ability to promote diversity and results in increased variation in terms of accuracy. 

\subsubsection{The GA-based incremental process vs Random Selection}
As stated in the section (\ref{approach}), the incremental GA-based process considers various types of scores to construct the exemplar set, including accuracy and loss. One of these types must be filled in as input to the algorithm. In order to check the impact of the score selection on the construction of the exemplar set and consequently on the model performance, we compare, in a first experiment, the incremental processes based on GA using accuracy and GA using loss \footnote{GA using accuracy denotes the variant of the genetic algorithm when we use the accuracy as a score in the fitness function, while GA using loss to denote that we use the loss measure in the fitness function} to the random selection process.

\begin{figure}
     \centering
     \begin{subfigure}[b]{0.45\textwidth}
         \centering
         \includegraphics[width=\textwidth]{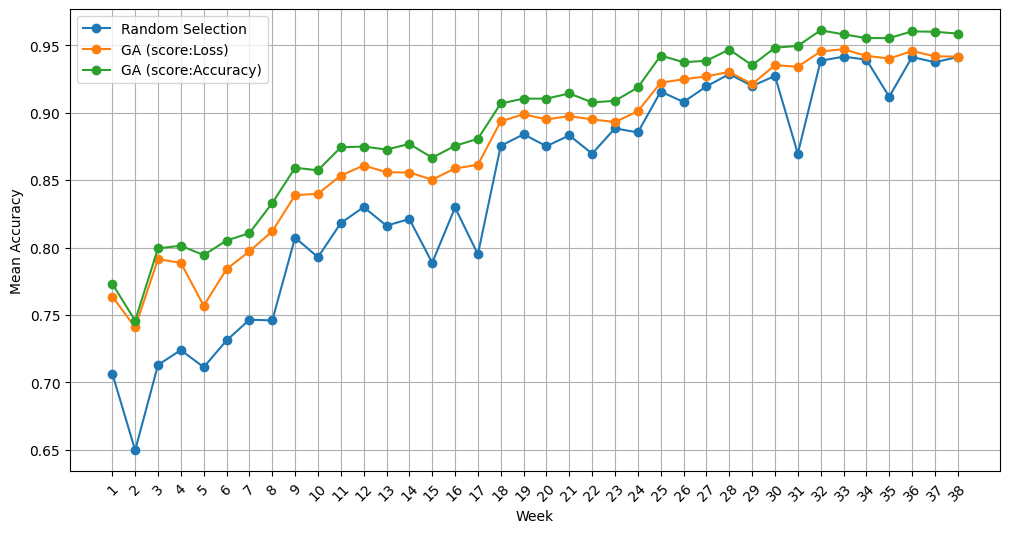}
         \caption{Mean Accuracy over 10 runs per week: exemplar size=80}
         \label{fig-1}
     \end{subfigure}
     \hfill
     \begin{subfigure}[b]{0.45\textwidth}
         \centering
         \includegraphics[width=\textwidth]{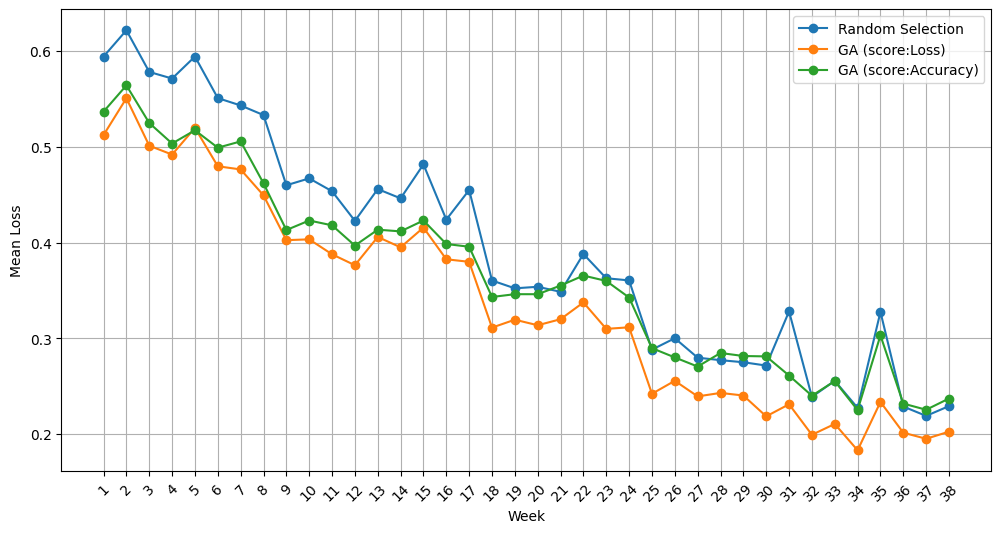}
         \caption{Mean Loss over 10 runs per week: exemplar size=80}
         \label{fig:three}
     \end{subfigure}
     \caption{GA (score=Loss) vs GA (score=Accuracy) vs Random Selection}
     \label{fig_}
\end{figure}

As shown in both Fig.\ref{fig-1} and Fig.\ref{fig:three}, our approach based on the use of GA either with a score type as accuracy or loss outperforms the existing random selection approach proposed in \cite{labba2022} in terms of both accuracy and loss. Indeed, our approach demonstrates a significant improvement, with up to a 10\% increase in accuracy and a noteworthy reduction in log loss of (-0.1) when compared to the random approach in \cite{labba2022}.

When assessing the various Genetic Algorithm (GA) variants by considering both accuracy and loss scores, a notable observation emerges. While the accuracy-based evaluation undoubtedly yields the best overall performance in terms of accuracy, the same cannot be said for the corresponding loss metric. Indeed, in our dataset, we have a total of 363 students, of whom 266 are classified as "pass" and 97 as "fail". It's important to note that in cases of data imbalance, accuracy alone may not provide an adequate measure to assess model performance, as it may be biased in favor of the dominant class-label. When generating the initial population in the GA approach, we make sure we maintain a certain balance between the class labels. However, during the mutation process, this condition can be violated, as this action is carried out randomly and on the basis of a given mutation probability. This may result in a good model performance that reaches only the dominant class. Whereas, when employing a loss-based score for the Genetic Algorithm (GA) variant, the model demonstrates superior results in terms of minimizing loss, although with a slight compromise in accuracy. We believe it's more effective to consider loss as a score for selecting the best training samples, as this reduces the overall loss while maintaining an acceptable measure of accuracy.

\section{Conclusion} \label{conclusion}

As data become available over time, traditional offline approaches of training and evaluating analytical models to predict students performance become obsolete and unsuitable. Nowadays, online incremental learning is increasingly being used to update online Machine Learning (ML) models with new data received over time. This work is concerned  with memory-based approaches that consist in using rehearsal techniques to recall a small training exemplar set that contains previous data and new data to retrain the online model. One of the major concerns in this regard is how to construct this training exemplar while receiving new data over time. Typically, a random selection of samples is made, which can deteriorate the model's performance. In this paper, we proposed a memory-based online incremental learning approach that is based on the use of the genetic algorithm heuristic to build the training exemplar set. The approach  respects the memory space constraints as well as the balance of class labels when forming the training exemplar.

Indeed, compared to an exiting approach based on random selection of training samples when building the training exemplar, our approach based on GA enhances the model accuracy up to 10\%. Further it shows a better stability and less variations in terms of accuracy. As a future work, we intend to evaluate the proposed approach with a variety of ML models in addition to random forest. Further, we intend to assess its effectiveness using other score types such as the F1-score. 

\bibliographystyle{apalike}

\bibliography{sample-base}

\end{document}